\documentclass[preprint,showpacs,showkeywords,preprintnumbers,amsmath,amssymb,superscriptaddress]{revtex4}
\usepackage{mathrsfs}
\usepackage{mathrsfs}
\usepackage{graphicx}
\usepackage{dcolumn}
\usepackage{bm}
\begin{document}

\title{Structured noise induced non-recross barrier escaping}

\author{Chun-Yang Wang}
\thanks{Corresponding author. Email: wchy@mail.bnu.edu.cn}
\affiliation{Shandong Provincial Key Laboratory of Laser Polarization and Information Technology, College of Physics and Engineering, Qufu Normal University, Qufu 273165, China}
\affiliation{State Key Laboratory of Theoretical Physics, Institute of Theoretical Physics, Chinese Academy of Sciences, Beijing 100190, China}


\begin{abstract}
The time-dependent barrier passage of a particle driven by the structured noise is studied in the field of a metastable potential.
Quantities such as the probability of passing over the saddle point and transmission coefficient of the escaping rate are calculated for a thimbleful of insight into the diffusion dynamical properties.
Results show that the barrier recrossing behavior is greatly reduced by the structured noisy environment.
Particles diffusion in such an dissipative environment tends to successfully escape from the potential well without any embarrassments.
\end{abstract}


\maketitle

\section{Introduction}\label{sec1}

Substantial progresses have been made during the last few years in the study of anomalous diffusions \cite{ad1,ad2,ad3} among which the barrier escaping problems are of fundamental interest.
A great amount of realistic events can be modeled by a single barrier escaping process within the framework of standard Brownian motion \cite{jc1,jc2,jc3}.
For an example, in the fusion of massive nuclei, particles are always assumed to escape from a metastable potential well.
The rate of escaping reveals then whether and to what extent the fusion process is successful.
However, the particles which have already escaped from the well may have some probability to recross the transition state.
This an inevitable problem in many cases which will significantly reduce the escaping rate.
Therefore the inhibition of barrier recrossing is of great importance for the shaping of compound nucleus.
The succeeding question is then how to realize the purpose of an effective inhibition.

Recently, a particular type of noisy environment namely the structured bath is advised which may provide us a convenient shortcut.
Obtaining from solving a second order stochastic differential equation driven by the Gaussian white noise,
three kinds of thermal colored noises (namely structured noises) are defined knowing as harmonic noise (HN), harmonic velocity noise (HVN) \cite{hvn} and harmonic acceleration noise (HAN) \cite{han}, respectively.
Study has found that HVN type structured noise can result in zero effective frequency-dependent friction.
And this leads to ballistic behaviors for the diffusion of free particles.
But it is not very clear for what will happen if the structured noises are used to drive a thermodynamically diffusion system or a dynamical process such as the fusion of massive nuclei.

Therefore, basing on these considerations, we present in this paper an elementary study concerning on the dynamics of barrier escaping in the structured noisy environment.
The paper is organized as follows:
In Sec. \ref{Sec2}, characteristic quantities such as the probability of passing over the saddle point for the description of a barrier passage are computed by analytically solving the generalized Langevin equation (GLE) of the system.
In Sec. \ref{Sec3}, dynamical details of the barrier escaping process are revealed by further computation on the transmission coefficient and Kramers rate in the neighbourhood of the saddle point of a metastable potential wherein this part the phenomenon of barrier recrossing is paid close attention.
Sec. \ref{Sec4} serves as a summary of our conclusion where some further discussions on the implicit application of this work are also made.

\section{Computation of the characteristic quantities}\label{Sec2}

Let us begin from the Mori-Kubo generalized Langevin equation (GLE) \cite{fdt1,ham} which can be derived from the Hamiltonian function by using of the Zwanzig-Mori projection method \cite{ZM1,ZM2} or the system-plus-reservoir model \cite{SPB1,SPB2,SPB3}:
\begin{equation}
m\ddot{x}(t)+\int_{0}^{\infty}dt'\beta(t-t')\dot{x}(t')+\partial_{x}
U(x)=\xi(t),\label{eq,GLE}
\end{equation}
where $\beta(t)$ is the friction kernel due to the structured noise relating via Kubo's second kind of fluctuation-dissipation theorem (FDT II) $\langle\xi(t)\xi(t')\rangle=mk_{B}T\beta(t-t')$ to the correlation function of each noise \cite{fdt2}
\begin{subequations}\begin{eqnarray}
&&\langle\xi(t)\xi(t')\rangle_{\textrm{HN}}=\frac{\eta}{\mu_{1}^{2}-\mu_{2}^{2}}
\left(\frac{1}{\mu_{1}}e^{\mu_{1}|t-t'|}-\frac{1}{\mu_{2}}e^{\mu_{2}|t-t'|}\right);\\
&&\langle\xi(t)\xi(t')\rangle_{\textrm{HVN}}=\frac{\eta}{\mu_{1}^{2}-\mu_{2}^{2}}
\left(-\mu_{1}e^{\mu_{1}|t-t'|}+\mu_{2}e^{\mu_{2}|t-t'|}\right);\\
&&\langle\xi(t)\xi(t')\rangle_{\textrm{HAN}}=\eta\delta(t-t')-\langle\xi(t)\xi(t')
\rangle_{\textrm{HN}}-\left(1-\frac{2\Omega^{2}}{\Gamma^{2}}\right)\langle\xi(t)\xi(t')\rangle_{\textrm{HVN}},
\end{eqnarray}\label{corf}\end{subequations}
here $\eta$ denotes the intensity of a Gaussian white noise, $\Gamma$ and $\Omega$ are the damping and frequency parameters in a RLC electric circuit respectively.
$\mu_{1}$ and $\mu_{2}$ are the roots of the characteristic equation $\mu^{2}+\Gamma\mu+\Omega^{2}=0$.
Symbol $m$ is the mass of a diffusing particle, $k_{B}$ the Boltzmann constant and $T$ the system temperature.

$U(x)$ in Eq.(\ref{eq,GLE}) is the external field which is always supposed to be metastable.
Mathematically, $U(x)=\frac{1}{2}m\omega^{2}_{a}(x-x_{a})^{2}$ at the ground state and $U(x)=U_{b}-\frac{1}{2}m\omega^{2}_{b}(x-x_{b})^{2}$ around the saddle point with barrier height $U_{b}$ and $\omega_{i}$ $(i=a,b)$ the corresponding frequencies.
Whatever form of the potential, the kinematics equation of the diffusion particle can be obtained from analytically solving the GLE.
For an example, near the saddle point of a metastable potential,
\begin{eqnarray}
x(t)=\langle x(t)\rangle+\int^{t}_{0}H(t-t')\xi(t')dt'\label{eq,x}
\end{eqnarray}
where $\langle x(t)\rangle=\left[1+\omega^{2}_{b}\int^{t}_{0}H(t')dt'\right]x_{0}+H(t)v_{0}$ is the mean displacement of the particle and
$H(t)=\mathscr{L}^{-1}[s^{2}+s\hat{\beta}(s)-\omega^{2}_{b}]^{-1}$ the namely response function \cite{rest}.

With these results, the probability of passing over the saddle point can be obtained from a simple integration on
the reduced distribution
\begin{eqnarray}
W(x,t; x_{0},v_{0})=\frac{1}{\sqrt{2\pi\sigma_{x}^{2}(t)}}\textrm{exp}[{-\frac{(x-\langle
x(t)\rangle)^{2}}{2\sigma^{2}_{x}(t)}}].
\end{eqnarray}
That is
\begin{eqnarray}
&&\chi(x_{0},v_{0};t)=\int^{\infty}_{0}W(x,t; x_{0},v_{0})dx\nonumber\\&&
\hspace{2cm}=\frac{1}{2}\textrm{erfc}\left(-\frac{\langle x(t)\rangle}{\sqrt{2}\sigma_{x}(t)}\right),\label{ksai}
\end{eqnarray}
which is generally a supplementary error function where
\begin{eqnarray}
\sigma^{2}_{x}(t)=\int^{t}_{0}dt_{1}H(t-t_{1})\int^{t_{1}}_{0}dt_{2}\langle \xi(t_{1})\xi(t_{2})\rangle H(t-t_{2})
\end{eqnarray}
is the variance of $x(t)$ resulted also from Laplacian solving the GLE.

\section{Barrier escaping and rescrossing} \label{Sec3}

The key point to investigate the barrier escaping dynamics is to obtain the escaping rate and relevant quantities.
Instead of directly solving the relevant Fokker-Planck equation \cite{SPB1},
we make the derivations from a newly developed method namely reactive flux (RF) \cite{rf1,rf2,jcp} which is base on the Langevin trajectory dynamics.
Mathematically the rate is defined as
\begin{eqnarray}
k(t)&=&\frac{m}{Qh}\int^{\infty}_{-\infty}dx_{0}
\int^{\infty}_{-\infty}v_{0}
W_{\textrm{st}}(x_{0},v_{0})\delta(x_{0}-x_{b})\chi(
x_{0},v_{0};t)dv_{0},
\label{kt}
\end{eqnarray}
where $W_{\textrm{st}}(x_{0},v_{0})$ is an initial Boltzmann form stationary probability distribution and $Q$ its partition function.
$h$ is a cell constant in the phase space.
In the spirit of RF method, initial conditions are assumed to be at the top of the barrier.
By investigating the ensemble of trajectories starting with identical initial conditions but experiencing different stochastic histories, a particle is determined for whether has successfully escaped or not.

Generally, for the convenience of computation, the rate can be written as a traditional transition state theory (TST) rate $k^{\textrm{TST}}=\frac{k_{B}T}{Qh}e^{-U_{b}/k_{B}T}$ \cite{TST1,TST2,TST3} multiplied by a transmission coefficient $\kappa(t)$, i.e.
\begin{eqnarray}
k(t)&=&k^{\textrm{TST}}\frac{m}{k_{B}T}\int^{\infty}_{-\infty}v_{0}
\textrm{\textrm{exp}}\left(-\frac{mv^{2}_{0}}{2k_{B}T}\right)
\chi(x_{0}=x_{b},v_{0};t)dv_{0},\label{eq,rate}
\end{eqnarray}
where
$W_{\textrm{st}}(x_{0},v_{0})=\textrm{exp}[-(mv^{2}_{0}+m\omega^{2}_{b}x^{2}_{0})/2k_{B}T]$
is determined by the initial conditions \cite{adel}.
Then it leads to
\begin{eqnarray}
\kappa(t)=\left(1+\frac{m\sigma^{2}_{x}(t)}{k_{B}TH^{2}(t)}\right)^{-1/2}.\label{eq,kappa}
\end{eqnarray}
This very formula of fractional reactive index $\kappa(t)$ is identical to the previous results \cite{kapp} and
leads immediately to the Kramers formula for the rate constant which is derived from the namely flux-over-population algorithm \cite{fop,kramers}.

\begin{figure}
\includegraphics{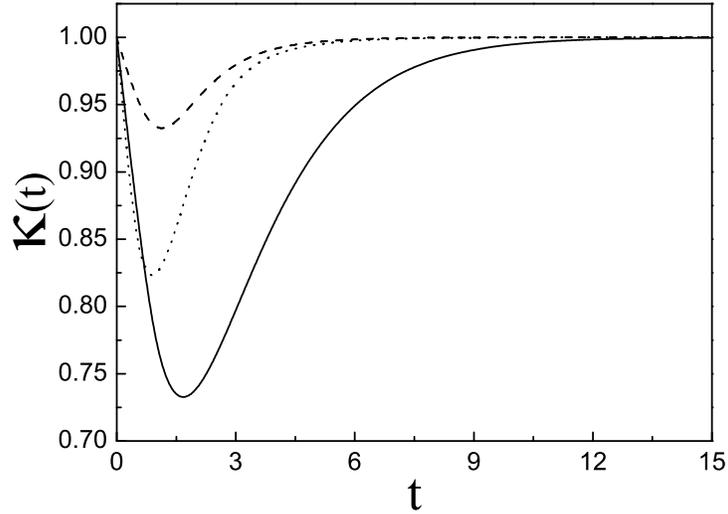}
\caption{Instantaneous $\kappa(t)$ for various kind of structured-noise environments: HN (solid), HVN (dash) and HAN (dot).
Parameters used are: $m=k_{B}T=\Gamma=\eta=\omega_{b}=1.0$ and $\Omega^{2}=2.0$ as well as $x_{0}=-0.5$ and $v_{0}=1.0$.}\label{Fig1}
\end{figure}

\begin{figure}
\includegraphics{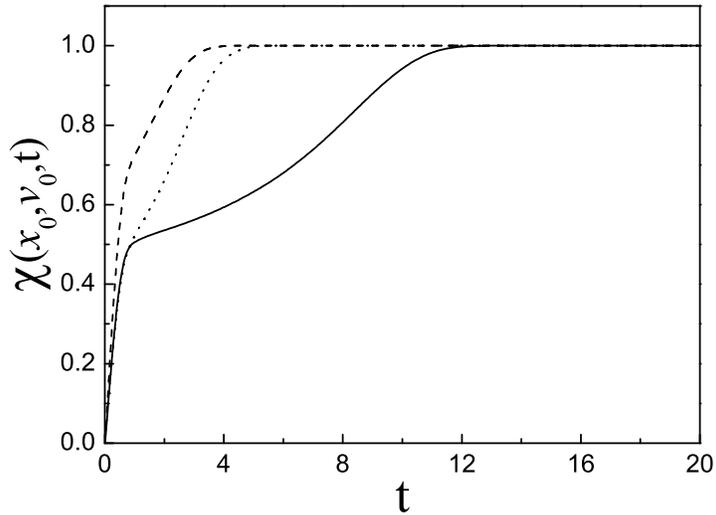}
\caption{Time-dependent barrier passing probability $\chi(x_{0},v_{0};t)$ for various kind of structured-noise environments:
HN (solid), HVN (dash) and HAN (dot). Identical dimensionless parameters are used as those shown in Fig.\ref{Fig1}.}\label{Fig2}
\end{figure}

In the following calculations, we rescale all the quantities so that dimensionless units such as $k_{B}T=1.0$ can be used.
Firstly, in Fig.\ref{Fig1} the transmission coefficient $\kappa(t)$ is plotted as a function of time $t$ for various kind of structured noisy environments.
From which we can see that $\kappa(t)$ quickly approaches constant 1 after a short period of oscillation.
Remembering that $\kappa(t)$ means the probability for a particle that has already escaped from the metastable well to recross the barrier.
Therefore this reveals that there is no recrossing in the structured noisy barrier escaping.
This is nontrivial in comparison with the non-Ohmic damping cases where a certain amount of barrier recrossing is always met \cite{jcp,kapp}.

For a better understanding of this result, we plot in Fig.\ref{Fig2} the corresponding barrier passing probability
$\chi(x_{0},v_{0};t)$ for various kinds of structured noisy environments.
From which we can see that, $\chi( x_{0},v_{0};t)$ increases also quickly to 1 as the time goes on.
This reveals that almost all the particles can escape from the potential well without any embarrassment.
This is very different from the usual case.
The physical background for these non-trivial results may lie in that structured noises are generally believed to originate from the velocity-dependent coupling between the system and heat bath \cite{vvc,cpl}.
They are in reality quasi-monochromatic noises and will reduce to Gaussian white noise if some prerequisite conditions are satisfied.
This assorts the structured noises a particular group of noise that is neither trivial as the white noise nor special as the non-Ohmic noise.
Therefore, the structured noisy environment may in some case lead to an anomalous dissipative system with zero effective friction.

\section{Summary and discussion} \label{Sec4}

In summary we have studied in this paper the dynamics of barrier escaping in the structured noisy environment.
Due to the zero effective friction caused by the quasi-monochromatic properties of the structured noises,
no barrier recrossing is witnessed.
The diffusion particles can escape from a metastable potential well without any embarrassment.
This may inspire us to perform much further studies concerning on the usage of structured noisy environments in particular in the experimental research of some realistic and important field such as the fusion of massive nuclei.

Structured noises and correspondingly defined structured heat baths are a group of newly suggested dissipative environments which has abundant of physical scenarios.
The most obvious examples can be found in the cases such as a charged particle interacting via dipole coupling with a black-body radiation field in Josephson junctions \cite{bjd,bbf} and in some electro-magnetic problems such as super-conducting quantum interference device \cite{spb4,SPB2}, where
the basic variable, i.e. the magnetic flux, is coupled to a quantity with the dimension of electric current.
Therefore, we believe the present study, although is not very penetrated into all the secrets, will in no doubt provide useful information for better understanding the relevant problems of these fields.

\section * {ACKNOWLEDGEMENTS}

This work was supported by the Shandong Province Science Foundation for Youths under the Grant No.ZR2011AQ016, the Open Project Program of State Key
Laboratory of Theoretical Physics, Institute of Theoretical Physics, Chinese Academy of Sciences, China under the Grant No.Y4KF151CJ1.

\end{document}